\documentclass[aps,prd,preprint,tightenlines,superscriptaddress,nofootinbib]{revtex4}
\usepackage{epsfig}
\usepackage{graphicx}
\usepackage{psfrag}
\usepackage{amsmath,amssymb}
\usepackage{colordvi}
\usepackage{amsfonts}
\usepackage{enumerate}
\usepackage{slashed}
\usepackage{color}
\usepackage{xcolor}

\newcommand{\olra}{\overleftrightarrow}

\begin{document}

\title{Proton Spin Structure from Measurable Parton Distributions}
\author{Xiangdong Ji}
\affiliation{Department of Physics, Shanghai Jiao Tong University, and
Shanghai Key Lab for Particle Physics and Cosmology, Shanghai, 200240, P. R. China}
\affiliation{Center for High-Energy Physics, Peking University, Beijing, 100080, P. R. China}
\affiliation{Maryland Center for Fundamental Physics, University of Maryland, College Park, Maryland 20742, USA}
\author{Xiaonu Xiong}
\affiliation{Center for High-Energy Physics, Peking University, Beijing, 100080, P. R. China}
\affiliation{Nuclear Science Division, Lawrence Berkeley
National Laboratory, Berkeley, CA 94720, USA}
\author{Feng Yuan}
\affiliation{Nuclear Science Division, Lawrence Berkeley
National Laboratory, Berkeley, CA 94720, USA}
\date{\today}
\vspace{0.5in}
\begin{abstract}
We present a systematic study of the proton spin structure
in terms of measurable parton distributions. For a transversely-polarized
proton, we derive a polarization sum rule from the leading
generalized parton distributions appearing in hard exclusive processes.
For a longitudinally-polarized proton, we obtain a helicity decomposition from
well-known quark and gluon helicity distributions and orbital angular-momentum
contributions. The latter is shown to be related to measurable subleading
generalized parton distributions and quantum-phase space Wigner distributions.
\end{abstract}

\maketitle

{\it 1. Introduction.}
Understanding the internal structure of the proton, its spin structure in particular,
has been a driving motive for intense activities in hadron physics in the last two decades.
Great progresses have been made from both experiment and theory sides.
Studies of deep-inelastic scattering (DIS) and related hard processes at the electron
facilities at SLAC, DESY, CERN, and Jefferson Lab, and of polarized
proton-proton collisions at relativistic heavy-ion collider (RHIC),
have generated a large body of experimental data, revealing
delicate roles of quarks and gluons in the proton spin.
These developments have stimulated theoretical advances from
simple parton model description of the nucleon structure to
multi-dimension distributions of partons, including
the generalized parton distributions (GPD's), the transverse momentum dependent
parton distributions (TMD's), and the quantum phase space Wigner distributions.
Together with the advances made in the lattice quantum chromodynamics (QCD),
these developments have provided us not only deep insights for the partonic
structure of the nucleon, but also the great opportunities to study the strong
interaction physics, such as the QCD factorization for hard processes, and the
universality of the associated parton distributions. A recent summary on
the experimental and theoretical status can be found in~\cite{Boer:2011fh}.

One of the key developments in understanding the spin structure of the proton
is the spin sum rule derived by one of the authors~\cite{Ji:1996ek},
where the total contributions to the spin from the quark
and gluons can be measured through their GPDs separately~\cite{Ji:1996nm}.
The partonic interpretation of this spin sum rule is, however, obscure.
In particular, for a transversely polarized proton, there appear conflicting partonic
interpretations of the spin~\cite{Burkardt:2002hr,Leader:2011cr}.
For longitudinal polarization, one can in principle deduce the quark orbital angular
momentum (OAM) by subtracting the quark helicity distribution.
However, it has not been able to identify a direct probe for the quark OAM
in physical processes. Meanwhile, the relation between the gauge-invariant quark OAM
and the canonical OAM~\cite{Jaffe:1989jz,Bashinsky:1998if} has been a confusing
issue in formulating a helicity sum rule with simple physical significance.

In this paper, we will address the above important questions by
systematically seeking a partonic interpretation of the proton spin and the
experimental measurability of the relevant distributions. 
We explain why a simple partonic
sum rule exists only for the transverse polarization. We find that
the gauge-invariant OAM contribution to the proton helicity
is related to twist-two and three GPD's which are measurable in hard exclusive processes.
Finally, the canonical OAM distribution in the
light-cone gauge is related to a Wigner distribution~\cite{Lorce:2011kd,Hatta:2011ku}, 
which is accessible through
certain hard processes. Our discussions are mainly focused on quarks,
but they can be easily extended to gluons.

Our starting point is the matrix element of the QCD AM density $M^{\mu\alpha\beta}$
in the nucleon plane-wave state \cite{Jaffe:1989jz}
\begin{eqnarray}
 && ~~~\langle PS|\int d^4\xi  M^{\mu\alpha\beta}(\xi)  |PS\rangle  = J \frac{2S_\rho P_\sigma}{M^2}
 {{(2\pi)^4
  \delta^4(0)}} \nonumber \\
 && \times
  \left(\epsilon^{\alpha\beta\rho\sigma}P^\mu + \epsilon^{[\alpha\mu\rho\sigma}P^{\beta]} -({\rm trace})\right)+ \cdots \ , \label{energy}
\end{eqnarray}
where $\xi^\mu$ is the space-time coordinates, $P^\mu$ and $S^\mu$  ($S\cdot P=0$, $S^2=-M^2$) 
are the four-momentum and polarization of the nucleon,
and $J=1/2$ and $M$ are the spin and mass, respectively. 
The $[\alpha\dots\beta] $ indicates antisymmetrization of the two indices.
The above equation is fully Lorentz-covariant and can be specialized to any
frame of references. To seek the partonic interpretation, we consider the nucleon in
the Infinite Momentum Frame(IMF) along the z-direction and take $\mu$ to 
be + component [$P^+=(P^0+P^3)/\sqrt{2}$].
Because of the antisymmetry between $\alpha$ and $\beta$, 
the leading component of the angular momentum density
comes from $\alpha=+$ and $\beta=\perp=(1,2)$. This is only possible if the nucleon is transversely
polarized ($S_\perp$) and the matrix element reduces to
\begin{equation}
\langle PS | \int d^4\xi M^{++\perp}|PS\rangle = J \left[\frac{3(P^+)^2S^{\perp'}}{M^2}\right]{{(2\pi)^4
  \delta^4(0)}} \  ,\label{transverse}
\end{equation}
where $S^{\perp'}=\epsilon^{-+\perp\rho}S_{\rho}$ with convention of $\epsilon^{0123}=1$.
In the above equation, a factor of 2 comes from the first term in the bracket of Eq.~(\ref{energy}),
whereas the second term contributes to a factor 1 because of the antisymmetric feature of indices 
$\alpha$ and $\beta$.

The longitudinal polarization supports the matrix element of
the next-to-leading AM tensor component $M^{+12}$,
\begin{equation}
\langle PS | \int d^4\vec{\xi} M^{+12}|PS\rangle = J (2S^+) {{(2\pi)^4\delta^4(0)}}\ , \label{longitudinal}
\end{equation}
which has one $P^+$-factor less. Thus the nucleon helicity $J$ is a subleading light-cone
quantity, and a partonic interpretation will in general involve parton transverse-momentum and
correlations.

The above result is in contrary to the common intuition about the role of spin-1/2 particle
polarization in hard scattering processes: The polarization vector $S^\mu$ has the leading
light-cone component $S^+=P^+$ when the nucleon is longitudinally polarized, and the
transverse component $S^\perp$ is subleading in the IMF.

{\it 2. Transverse-polarization Sum Rule.}
According to Eq.~(\ref{transverse}), one expects a simple partonic interpretation of the
transverse proton polarization from the leading parton distributions. Indeed, the quark AM
sum-rule derived in terms of the quark distribution $q(x)$ and
GPD $E(x, 0, 0)$ is exactly of this type~\cite{Ji:1996ek},
\begin{equation}
   J_q = \frac{1}{2} \sum_i \int dx x\left[q_i (x) + E_i(x, 0, 0)\right] \ ,\label{Jisumrule}
\end{equation}
where $i$ sums over different flavor of quarks, and similarly for the gluon AM.
We emphasize that this spin sum rule is frame-independent.
In Ref.~\cite{Burkardt:2002hr}, Burkardt has proposed an interesting
explanation of the above result in the impact parameter space, in which a transversely polarized
nucleon state fixed in the transverse plane generates a spatial asymmetric parton density
$q(x, b_\perp)$, which yields to the parton's AM contribution to the transverse spin. Note that
the above sum rule is different from that of E. Leader \cite{Leader:2011cr}, because the transverse
angular $\vec{J}_{\perp}$ does not commute with the Lorentz boost along the z-direction.

To attribute the above sum rule with a simple parton picture,
one has to justify that $(x/2)(q(x)+E(x))$ is the
transverse AM density in $x$, i.e., it is just the contribution to the transverse
nucleon spin from partons with longitudinal momentum $xP^+$.
This can be done easily.
Define the quark longitudinal momentum
density $\rho^{+}(x, \xi,S^\perp)$ through
\begin{equation}
   \rho^{+} (x, \xi,S^\perp) = x\int \frac{d\lambda}{4\pi} e^{i\lambda x}
     \langle PS^\perp|\overline \psi(-\frac{\lambda n}{2},\xi)\gamma^+\psi(\frac{\lambda n}{2},\xi)|PS^\perp\rangle \ ,
\end{equation}
where $n$ is the conjugation vector associated with $P$:
$n=(0^+,n^-,0_\perp)$ with $n\cdot P=1$. A careful calculation shows that beside the usual momentum distribution,
it has an additional term
\begin{equation}
   \rho^{+} (x, \xi,S^\perp)/P^+=  xq(x)+\frac{1}{2}x\left(q(x)+E(x)\right)\lim_{\Delta_\perp\rightarrow 0}\frac{S^{\perp'}}{M^2} \partial^{\perp_\xi} e^{i\xi_\perp\Delta_\perp}
   \end{equation}
where the $\xi^\perp$-dependence comes from the slightly off-forward matrix element, which acts like a ``distribution" in mathematical sense:
vanishing normally but non-zero when integrated with some kernels. The parton contribution to the transverse polarization 
is just the transverse-space moment of
$\rho^{+}(x, \xi,S^\perp)$,
\begin{equation}
    S_\perp^q(x) = \frac{M^2}{2P^+S^{\perp'}(2\pi)^2\delta^2(0)}\int d^2\xi \xi^{\perp} \rho^{+}(x, \xi,S^\perp) = \frac{x}{2}(q(x)+E(x)) \ . \label{jx}
\end{equation}
where we have included the contribution from the energy-momentum component $T^{+\perp}$ through Lorentz symmetry.

{\it 3. Helicity Sum Rule.}
Most of the experimental probes on the nucleon spin use the longitudinal
polarization, and thus it is natural to explore the nucleon
helicity in parton picture.
Considering the z-component of the quark AM, we have,
\begin{eqnarray}
  &\!\!&J^3 = \int d^3\vec{\xi} M^{+12}(\xi) \nonumber \\
  \!\!&=\!\!& \int d^3\vec{\xi} \left[\overline{\psi}
 \gamma^+(\frac{\Sigma^3}{2})\psi + \overline{\psi}\gamma^+ \left(\xi^1 (iD^2) - {{\xi^2 (iD^1)}}\right) \psi
 \right] \ .
\label{quarkang}
\end{eqnarray}
while the quark helicity is well-known to have a simple parton density interpretation. However,
the quark OAM involves transverse component of the gluon field, and thus is related to
three-parton correlations. We notice recent 
research activities aiming at different 
decompositions of the nucleon spin~\cite{Chen:2008ag}, which we will not address
in this paper (see also a recent comment on these developments~\cite{Ji:2012gc}).

Thus a partonic picture of the orbital contribution to the nucleon helicity
{\it necessarily involves parton's transverse momentum}. In other words,
TMD parton distributions are the right
objects for physical measurements and interpretation.
In recent years, TMD's and novel effects associated with them have been
explored extensively in both theory and experiment~\cite{Boer:2011fh}.
An important theoretical issue related to them is gauge invariance.
Whenever a canonical momentum of color-charged particles appears,
the gauge symmetry requires that the gauge potential $A^\mu$ must be present
simultaneously. This is already true when parton's longitudinal momentum
distribution is considered: In factorization theorems for DIS,
the physical parton represents a gauge-invariant object with a
gauge link extended from the location of parton field to infinity
along the conjugating light-cone direction $n^\mu$,
\begin{equation}
        \Psi_{LC}(\xi)= {{P}}\left[\exp\left(-ig\int^\infty_0 d\lambda n \cdot A(\lambda n+\xi)\right)\right]   \psi(\xi) \ .\label{lc}
\end{equation}
where $P$ indicates path ordering.
Therefore, in perturbative diagrams, a parton with
momentum $k^+ = xP^+$ represents in fact the sum of all diagrams with longitudinal gluons
involved.

When considering parton's transverse momentum, we also need
appropriate gauge links formed of gauge potentials.
The choice for the gauge links, however, is scattering-process
dependent~\cite{Collins:2002kn}. As a consequence, there is no unique definition for the TMD's.
In practical applications, two choices stand out. First one
uses the same light-cone gauge link as shown in the above.
This choice does lead to light-cone singularities which must be
addressed properly in actual calculations~\cite{Ji:2004wu}. The second choice is a straightline
gauge link along the direction of spacetime position $\xi^\mu$,
\begin{equation}
        \Psi_{FS}(\xi)= {{P}}\left[\exp\left(-ig\int^\infty_0 d\lambda \xi \cdot A(\lambda\xi)\right)\right] \psi(\xi) \ . \label{fs}
\end{equation}
The link reduces to unity in Fock-Schwinger gauge, $\xi\cdot A(\xi)=0$. 
The gauge invariant parton fields $\Psi(\xi)$ are defined in the IMF 
which is the basis of partonic interpretation.

To investigate parton's OAM contribution to the proton helicity,
one also needs their transverse coordinates. The most natural concept
is a phase-space Wigner distribution, which was first introduced in
Ref. \cite{Belitsky:2003nz}. A Wigner distribution operator
for quarks is defined as
\begin{equation}
   \hat {\cal W}(\vec{r},k) = \int \overline{\Psi}(\vec{r}-\xi/2)
    \gamma^+\Psi(\vec{r}+\xi/2) e^{ik\cdot \xi} d^4\xi \ ,
\end{equation}
where 
$\vec{r}$ is the quark phase-space position and $k$
the phase-space four-momentum,
and $\Psi$ follows the definitions of Eqs.~(\ref{lc},\ref{fs}). 
They represent the two different choices for the gauge links
associated with the quark distributions.
Including the gauge links in Eqs.~(\ref{lc},\ref{fs}) makes the above definition
gauge invariant. 
However, they do depend on the choice of the gauge link~\cite{Collins:2002kn},
as we will show below.
The Wigner distribution can be define as the expectation
value of $\hat {\cal W}$ in the
nucleon state,
\begin{eqnarray}
  &&  ~~   W (k^+=xP^+, \vec{b}_\perp, \vec{k}_\perp) \nonumber \\
                &=& \frac{1}{2} \int \frac{d^2\vec{q}_\perp}{(2\pi)^3} \int \frac{dk^-}{(2\pi)^3}
                    e^{-i\vec{q}_\perp\cdot\vec{b}_\perp}
                   \left \langle \frac{\vec{q}_\perp}{2}\left|
          \hat {\cal W}(0,k)\right|-\frac{\vec{q}_\perp}{2}\right\rangle \ . 
\end{eqnarray}
where the nucleon has definite helicity 1/2.
The quark's OAM distribution follows from the intuition,
\begin{equation}
  L(x) = \int  (\vec{b}_\perp \times \vec{k}_\perp) W(x, \vec{b}_\perp,  \vec{k}_\perp)  d^2\vec{b}_\perp d^2 \vec{k}_\perp \ ,
\end{equation}
from partons with longitudinal momentum $xP^+$.

For our purpose, the most appealing choice is $\Psi_{FS}$ because it leads to a light-cone
AM density both calculable on lattice and measurable experimentally.
To demonstrate this, we need the Taylor expansion,
\begin{eqnarray}
 \overline{\Psi}_{FS}(-\xi/2)\gamma^+ \Psi_{FS}(\xi/2)
=  \sum_{n=0}^\infty {\overline \psi}(0)
\gamma^+ {{\olra{D}}}^{\mu_1} ...  {{\olra{D}}}^{{\mu_n}} \psi(0) \xi_{\mu_1} ... \xi_{\mu_n} \ .
\end{eqnarray}
It follows that
\begin{eqnarray}
 &&  \int x^{n-1} L_{FS}(x)dx  =  \frac{1}{\langle PS|PS\rangle}\langle PS| \int d^3\vec{r}\sum_{i=0}^{n-1} \frac{1}{n}
 \overline{\psi}(\vec{r}) \nonumber\\ && \times
 (in\cdot D)^i  (\vec{r}_\perp \times i\vec{D}_\perp) (in\cdot D)^{n-1-i} \psi(\vec{r})
   |PS\rangle  \  .
\end{eqnarray}
The right-hand side is related to the matrix elements of twist-2 and twist-3 operators, which
are extractable from experimental data on twist-3 GPD's~\cite{gpd3,wignermeasure}.
Because there is no light-cone non-local operators involved, it can also be calculated
in lattice QCD~\cite{Musch:2010ka}. We emphasize that $L_{FS}(x)$ is not the same as the OAM density defined
through the generalized AM density in Ref. \cite{Hoodbhoy:1998yb}. The difference is a twist-three GPD contribution
proportional to the gluon field $F^{+\perp}$.

The total OAM sum rule in term of parton's Wigner distribution,
\begin{eqnarray}
\frac{\langle PS| \int d^3\vec{r} ~\overline{\psi}(\vec{r}) \gamma^+ (\vec{r}_\perp \times i\vec{D}_\perp) \psi(\vec{r})
   |PS\rangle}{\langle PS|PS\rangle}
    =  \int  (\vec{b}_\perp \times \vec{k}_\perp) W_{FS} (x, \vec{b}_\perp,  \vec{k}_\perp)  dx {d^2\vec{b}_\perp d^2 \vec{k}_\perp}
\label{sum}
\end{eqnarray}
which gives a parton picture for the gauge-invariant OAM \cite{Ji:1996ek}, although
the straightline gauge link destroys the straightforward parton density interpretation.

Other choices of gauge links yield different Wigner distributions and
hence different partonic OAM distributions $L(x)$.  However, so long the gauge
link between $[-\xi/2, \xi/2]$ is smoothly differentiable, Eq. (\ref{sum}) remains valid.
This is one of the important virtue of the gauge-invariant approach.
However, for partons with the light-cone gauge link, $\Psi_{LC}$, the above sum rule
is invalid, as we shall see below.

{\it 4. Canonical Orbital Angular Momentum.}  The quark contribution to the canonical orbital
angular momentum was explored in Ref.~\cite{Bashinsky:1998if},
\begin{eqnarray}
   l_q (x) &=& \frac{1}{(2\pi)^22P^+\delta^{2}(0)}\int \frac{d\lambda}{2\pi}  e^{ix\lambda} d^2\xi 
    \langle PS|\overline\psi(-\frac{\lambda n}{2},\xi) \gamma^+
    \nonumber \\
  &&   \times
   (\xi^1i\partial^2-\xi^2i\partial^1)\psi(\frac{\lambda n}{2},\xi)|PS\rangle \ .
  \label{bjparton}
\end{eqnarray}
This definition represents the canonical OAM in the light-cone gauge $A^+=0$, and
is not gauge invariant.
A gauge-dependent quantity is in principle not measurable experimentally.
However, sometimes one can fortunately find its
{\it gauge-invariant extension} (GIE) physically measurable. A GIE of a gauge-variant quantity is a fixed-gauge result
gauge-invariantly extrapolated to any other gauge.
A GIE of the partial derivative in $ A^+=0$ gauge is
\begin{equation}
   i\partial^\perp_\xi = iD^\perp_\xi + \int^{\xi^-} d\eta^-  L_{[\xi^-,\eta^-]} gF^{+\perp}(\eta^-,\xi_\perp) L_{[\eta^-,\xi^-]} \  ,
\end{equation}
which is uniquely defined, and $L_{[\xi^-,\eta^-]}$ is the light-cone gauge link connecting $\xi^-$ and $\eta^-$.
One can plug this into Eq.~(\ref{bjparton}) to obtain an GIE
of $l_q(x)$ away from $A^+=0$. The covariant derivative term is just what we have discussed
before. The second term involves a non-local operator along the light-cone, and has obscure physical
meaning other than in the light-cone gauge. Its matrix
element is in principle related to the twist-three GPD's~\cite{gpd3}, and an infinite number of moments
are involved due to non-locality. Therefore, we arrive at the interesting conclusion that
$l_q(x)$ in light-cone gauge is actually accessible through twist-two and three GPD's, which
is consistent with what Hatta has concluded recently~\cite{Hatta:2011ku}.

A clear parton picture emerges through connections between
$l_q(x)$ and TMD's and Wigner distributions~\cite{Lorce:2011kd,Hatta:2011ku}.
One can introduce a Wigner distribution with
the gauge link in the light-cone direction, $W_{LC}(x, \vec{b}_\perp,  \vec{k}_\perp)$.
Integration over the impact parameter space $\int d^2\vec{b}_\perp W_{LC}$ generates
quark-spin independent TMD's. It can be shown that the canonical AM
distribution in $A^+=0$ gauge as defined in~\cite{Bashinsky:1998if} can be obtained
from the simple moment of a gauge-invariant Wigner distribution,
\begin{equation}
  l_q(x) = \int  (\vec{b}_\perp \times \vec{k}_\perp) W_{LC}(x, \vec{b}_\perp,  \vec{k}_\perp)  {{d^2\vec{b}_\perp d^2 \vec{k}_\perp}} \ .
\end{equation}
From the discussion of the previous
paragraph, this also implies constraints on the moments of Wigner distributions from the GPD's.
Finally, the canonical OAM in light-cone gauge acquires the simple
parton sum rule in the quantum phase space~\cite{Lorce:2011kd,Hatta:2011ku},
\begin{eqnarray}
&&l_q = \frac{\langle PS| \int d^3\vec{r}\,\, \overline{\psi}(\vec{r}) \gamma^+(\vec{r}_\perp \times i\vec{\partial}_\perp) \psi(\vec{r})
   |PS\rangle}{\langle PS|PS\rangle}  \nonumber \\
    && =  \int  (\vec{b}_\perp \times \vec{k}_\perp) W_{LC}(x, \vec{b}_\perp,  \vec{k}_\perp)  dx {{d^2\vec{b}_\perp d^2 \vec{k}_\perp}} \ .
\end{eqnarray}
The measurability of this Wigner distribution will be studied in a future publication~\cite{wignermeasure}.

{\it 5. Conclusion.}
To summarize, we explored systematically parton pictures for the proton spin, and
achieved a number of important results. For the transverse polarization, we found that
it is simple to interpret in terms of parton AM density measurable through twist-two GPD's.
For the nucleon helicity, the gauge-invariant parton picture can be probed through
twist-two and three GPDs, and also calculable in lattice QCD.
A simpler parton picture in the light-cone gauge can be established through the quantum
phase sapce Wigner distribution, and can be measured through either twist-two and
three GPD's or directly from Wigner distribution. These results will stimulate further
theoretical developments and generate experimental interests to measure, particularly,
the parton OAM in hard scattering processes. Phenomenological studies
will be presented elsewhere.

We thank M. Burkardt and J.H.Zhang for helpful discussions. This work was partially supported by the U.
S. Department of Energy via grants DE-FG02-93ER-40762 and DE-AC02-05CH11231, and
a grant (No. 11DZ2260700) from the Office of Science and Technology in Shanghai Municipal Government .

\end{document}